\documentclass[nohyper,12pt,letterpaper]{JHEP}
\usepackage[dvips]{epsfig}
\usepackage{epsf,amsfonts,amssymb}

\newcommand{\be}{\begin{equation}}
\newcommand{\ee}{\end{equation}}
\newcommand{\ben}{\begin{displaymath}}
\newcommand{\een}{\end{displaymath}}
\newcommand{\bea}{\begin{eqnarray}}
\newcommand{\eea}{\end{eqnarray}}
\newcommand{\bean}{\begin{eqnarray*}}
\newcommand{\eean}{\end{eqnarray*}}
\newcommand{\nn}{\nonumber \\}
\newcommand{\ba}{\begin{array}}
\newcommand{\ea}{\end{array}}
\newcommand{\bi}{\begin{itemize}}
\newcommand{\ei}{\end{itemize}}

\def\G {\Gamma}

\def\e {\epsilon}

\newcommand\bfO{\mbox{\boldmath $\Omega$}}

\newcommand{\bftau}{\mbox{\boldmath $\tau$}}

\def\bfO{\mbox{\boldmath $\Omega$}}
\def\bfn{\mbox{\boldmath $\nabla$}}
\def\bftau{\mbox{\boldmath $\tau$}}
\def\bfsig{\mbox{\boldmath $\sigma$}}
\def\bfG{\mbox{\boldmath $\Gamma$}}
\def\bfD{\mbox{\boldmath ${\cal D}$}}

\input amssym.def
\input amssym.tex
\font\mybb=msbm10 at 10pt
\def\bb#1{\hbox{\mybb#1}}

\def\bR {\bb{R}}
\def\bE {\bb{E}}
\def\bI {\bb{I}}

\def\bC {\bb{C}}

\title{Sigma-model soliton intersections from exceptional
calibrations}

\author{R. Portugues and P. K. Townsend \\
E-mail: \email{R.Portugues, P.K.Townsend@damtp.cam.ac.uk}}

\abstract{A first-order `BPS' equation is obtained for 1/8 supersymmetric
intersections of soliton-membranes (lumps) of supersymmetric
(4+1)-dimensional massless sigma models, and a special
non-singular solution is found that preserves 1/4 supersymmetry. For
4-dimensional hyper-K\"ahler target spaces ($HK_4$) the BPS equation is
shown to be the low-energy limit of the equation for a Cayley-calibrated
4-surface in $\bE^4\times HK_4$. Similar first-order equations are found
for stationary intersections of Q-lump-membranes of the massive
sigma model, but now generic solutions preserve either 1/8 supersymmetry
or no supersymmetry, depending on the time orientation.}


\preprint{DAMTP-2002-35\\ \tt{hep-th/0203181}}

\begin{document}

\section{Introduction}

Supersymmetric sigma models, in dimension (1+1) or above, are
known to have a variety of supersymmetry-preserving, and hence
stable, soliton solutions, supported by a combination of topological and
Noether charges.  It is natural to consider the maximally-supersymmetric
models, with a total of 8 supersymmetries, because sigma models with fewer
supersymmetries can always be embedded in some model with maximal
supersymmetry, for which the target space is automatically hyper-K\"ahler
(HK), although not necessarily irreducible\footnote{e.g. a Calabi-Yau
3-fold ${\mathcal M}$ is a submanifold of the 12 real dimensional
target space ${\mathcal M}\times T^6$, which has holonomy
$SU(3)\subset Sp_3$, and therefore yields a sigma model with maximal
supersymmetry}.  In this context the fraction of supersymmetry preserved by a
supersymmetric solution may be\footnote{The  fraction 3/8 might be
possible, in principle, but no example is known.} 1/2, 1/4 or 1/8, and
this  fraction provides a convenient partial characterisation of the
solution; roughly speaking, the lower the fraction the more complicated
the solution is.

The basic sigma-model soliton is the 1/2 supersymmetric sigma-model lump
(see, for example, \cite{lump}), which  is a particle-like solution of
massless (2+1)-dimensional HK sigma models, but a  string-like solution
(the lump-string) of (3+1)-dimensional models, and so on up to a 3-brane
in the maximal dimension of (5+1). The sigma-model lump is `basic' in two
respects. Firstly, it gives rise to the 1/2 supersymmetric kink
soliton of the massive (1+1)-dimensional sigma model obtained by
non-trivial dimensional reduction on $S^1$ \cite{AT,tong}. Secondly, it is
the main ingredient in the construction of solutions preserving only 1/4
supersymmetry via `intersections'. The first example of this was
a non-singular 1/4-supersymmetric intersection of two kink domain walls in
massive (1+1)-dimensional HK sigma models of quaternionic
dimension $n\ge2$ \cite{GTT}. Another example, which is also possible for
$n=1$, is the kink-lump of massive (3+1) dimensional HK sigma
models \cite{GPTT}; this is a non-singular solution in which a lump-string
ends on a kink-domain-wall, thus imitating the physics of D2-branes in
type IIA superstring theory.

A IIA superstring ending on a D2-brane can be viewed, from an
11-dimensional point of view, as a pair of intersecting
supermembranes. Similarly, as we demonstrate here, the kink-lump of the
massive (3+1)-dimensional sigma model lifts to a pair of 1/4
supersymmetric intersecting lump-membranes of the massless
(4+1)-dimensional HK sigma model. However, our main interest in this
paper is intersections of these lump-membranes that preserve the
minimal fraction of 1/8 supersymmetry. We obtain a first-order `BPS'
equation for intersections of lump-membranes that is solved by our
explicit 1/4 supersymmetric configuration but for which the generic
solution preserves only 1/8 supersymmetry. Of course, there may be no
generic solution that satisfies the required boundary conditions in any
given model, but we propose a particular model that we believe will have
1/8 supersymmetric solutions of the desired type.

A trivial dimensional reduction of the massless (4+1)-dimensional HK sigma
model to a  massless (3+1)-dimensional model reduces the lump-membranes
to lump-strings. It also reduces the BPS equation for 1/8 supersymmetric
intersections of lump-membranes to a previously proposed BPS equation for
1/4 supersymmetric intersections of lump-strings \cite{NNS}. Another way
to interpret this fact is to say that any 1/8 supersymmetric
configuration of lump-membranes will degenerate to a 1/4 supersymmetric
solution in regions in which it becomes independent of one of the four
space coordinates, although this 1/4 supersymmetric solution is of an
essentially different type to the pair of intersecting membranes
mentioned above because it is a solution of the sigma-model equations
that is intrinsically 3-dimensional rather than 4-dimensional. If
this solution
becomes independent of {\it another} of the four space coordinates
then it reduces to the 1/2 supersymmetric, and intrinsically
2-dimensional, lump-membrane.

A {\it non-trivial} dimensional reduction of the massless
(4+1)-dimensional HK sigma model (when this is possible) yields a massive
(3+1)-dimensional HK sigma model, which admits static 1/2 supersymmetric
kink-membranes and stationary, charged, 1/4 supersymmetric
Q-lump-strings \cite{LA,GPTT}. Apart from the 1/4 supersymmetric kink-lump
mentioned above, and its stationary Q-kink-lump counterpart \cite{GPTT},
these massive models can also be expected to have 1/8 supersymmetric
intersections; for example, a BPS equation for 1/8 supersymmetric
intersecting Q-lump-strings is known
\cite{NNS}. If the HK target space is 4-dimensional, as it is for the
models we consider here, then any solution of the massive
(3+1)-dimensional sigma model is also a solution of the {\it same} massive
model in (4+1)-dimensions\footnote{This need not be true for
higher-dimensional HK target spaces because in this case the
(3+1)-dimensional model allows a more general potential than the
(4+1)-dimensional model; precisely this fact was exploited in
\cite{GTT}.}, but additional solutions are possible
in the higher dimension. In particular, our new BPS equation for 1/8
supersymmetric intersections of static lump-membranes of the massless
(4+1)-dimensional model extends to a pair of BPS equations for 1/8
supersymmetric intersections of stationary Q-lump-membranes. A curious
feature of generic solutions of these equations is that {\it supersymmetry
is preserved for only one time orientation}, despite the time-reversal
invariance of the full second-order field equations.

Supersymmetric sigma models in (5+1) dimensions with 4-dimensional HK
target spaces (quaternionic dimension 1) provide a low energy
effective description of the dynamics of an M5-brane in a spacetime of
the form $\bE^{(1,5)}\times HK_4 \times S^1$ in the special case of
vanishing worldvolume 3-form field strength and fixed position on $S^1$;
this truncation preserves the 8 supersymmetries that are
left unbroken by the combination of the M5-brane and the HK
background spacetime. Given that the
$\bE^{(1,5)}$ factor of the background is identified as the
fivebrane vacuum, the fivebrane dynamics is then governed by an action
for maps from
$\bE^{(1,5)}$ to $HK_4$, for which the low-energy limit is the
(1,0)-supersymmetric (5+1)-dimensional sigma model with $HK_4$ as its
target space (see \cite{pkt} for a review).

The 1/2 supersymmetric lump solitons of such
sigma models can be interpreted as smooth 3-brane intersections of
the original M5-brane with another M5-brane that is wrapped on a finite
area holomorphic 2-cycle of $HK_4$ \cite{BT,GPTT}. Actually, since the
intersection is smooth, one may consider the two intersecting M5-branes
in this background as a single K\"ahler-calibrated M5-brane \cite{GPT}.
In this M-theory context it is natural to suppose that intersections of
sigma-model solitons will also have an interpretation as calibrations.
However, whereas K\"ahler-calibrated surfaces are solutions of equations
that are homogeneous in derivatives, as are the BPS equations for
sigma-model solitons and their intersections, the equations of other
types of calibrated surfaces are {\it inhomogeneous in derivatives} and
therefore have no obvious interpretation as BPS equations in field theory.

Nevertheless, since a calibrated p-surface solves the equations of motion
for a p-brane, any solutions of the `linearized' calibration equation
(obtained by linearizing in derivatives) must solve the `linearized'
p-brane equations. In flat  space these `linearized' equations have no
interesting solutions, not surprisingly since the `linearized' equations
are truly linear in this case, being linear in fields as well as
derivatives. But in non-flat backgrounds the `linearized' calibration
equations will still be non-linear in fields and hence may have
interesting solutions.  In fact, we shall show that both the BPS equation
of \cite{NNS} for 1/4 supersymmetric intersecting sigma-model solitons
and the new BPS equation presented here for 1/8 supersymmetric
intersections are `linearizations' of the equations for, respectively,
associative and Cayley calibrations. Thus, intersecting HK sigma-model
solitons can  be understood in terms of exceptional calibrations
of a single M5-brane of M-theory. This result is reminiscent of
the interpretation of junctions of MQCD domain walls as Cayley calibrated
M5-branes \cite{GGT}, but we are not aware of any direct connection
to the results reported here.

\section{Supersymmetric HK sigma models}
\label{sec-two}

We begin with a brief discussion of the models to be considered. Let
$X^I$ ($I=1,\dots,4n$) be the scalar fields. Omitting fermions, the
massless supersymmetric sigma model in (5+1) dimensions has the action
\be\label{fiveone}
S_6[X]= {1\over 2} \int dt d^5x \left[ |\dot X|^2 -
|\nabla X|^2\right]\,,
\ee
where the overdot indicates differentiation with respect to the time
coordinate $t$, and $\nabla$ is the partial derivative with respect to
the five cartesian space coordinates. As the target space is
necessarily HK, there exist three complex structures
${\bf I}= (I_1,I_2,I_3)$, with matrix entries $(I_i)^I{}_J$ ($i=1,2,3$),
obeying the algebra of the quaternions
\be
I_iI_j = -\delta_{ij} \bI + \varepsilon_{ijk} I_k\, ,
\ee
where $\bI$ is the identity matrix. The triplet of complex structures
is associated with the triplet of (closed) Kahler 2-forms
\be
\bfO_{IJ} = g_{IK}{\bf I}^K{}_J\, .
\ee

The massless (5+1)-dimensional model can of course be trivially
dimensionally
reduced to any lower spacetime dimension, but given that the target space
admits a $U(1)$ Killing vector field (KVF) $k$, there is also the
possibility of a non-trivial dimensional reduction, achieved by setting
\be\label{redan}
\partial_5 X^I = m\, k^I\, ,
\ee
where $m$ is a mass-parameter. This reduction will preserve all 8
supersymmetries if $k$ is triholomorphic, which is the requirement
of vanishing ${\cal L}_k\bfO$, where ${\cal L}_k$ is the Lie derivative
with respect to the vector field $k$. The resulting maximally
supersymmetric (4+1)-dimensional sigma model has the action (again
supressing fermions)
\be
S_5[X]= {1\over 2} \int dt d^4x \left[ |\dot X|^2 - |\partial X|^2
- m^2|k|^2 \right],
\ee
where $\partial$ indicates differentiation with respect to the
{\it four} space coordinates. For $m\ne0$ there is a positive scalar
potential, given by the norm squared of the triholomorphic KVF $k$
\cite{AGF}; this potential vanishes at the isolated fixed points of
$k$.

Although many of our results on HK sigma models are valid for target
spaces of arbitrary quaternionic dimension $n$, we shall concentrate here
on the $n=1$ case because of its M-theory interpretation and the connection 
to
exceptional calibrations. Specifically, we shall consider 4-dimensional
HK manifolds with a triholomorphic $U(1)$ KVF.
In coordinates $(\varphi,{\bf X})$ for which this KVF is
\be
k = \partial/\partial \varphi\, ,
\ee
the metric is
\be\label{HK4}
ds^2 = U d{\bf X}\cdot d{\bf X} + U^{-1} (d\varphi +
{\bf A}\cdot d{\bf X})^2
\ee
where $U$ is a harmonic function on $\bE^3$ with isolated point
singularities and ${\bf A}$ is a 3-vector potential such that $\bfn \times
{\bf A}= \bfn \varphi$. If $\varphi$ is identified with period $2\pi$ then
completeness of the metric requires
\be
U = a + {1\over2}\sum_{k=1}^N |{\bf X}- {\bf X}_k|^{-1}
\ee
for some number $a$, integer $N$, and a choice of $N$
points ${\bf X}_k$ in $\bE^3$, which are
called the `centres' \cite{GH}. The triplet of  K\"ahler
2-forms is
\be
\bfO = {\cal D}\varphi d{\bf X} -{1\over2}U d{\bf X} \times d{\bf X}
\ee
where we have supressed the wedge product of forms, and defined the
covariant derivative
\be
{\cal D}\varphi = d\varphi + {\bf A}\cdot d{\bf X}\, .
\ee
The coefficients of  $\bfO$ are $\varphi$-independent, so $k$ is
triholomorphic.

For these 4-dimensional HK target spaces the massless (5+1)-dimensional
sigma model has the action
\be
S_6[\varphi,{\bf X}] = {1\over2}\int dtd^5x\, \left\{ U^{-1}\left[({\cal 
D}_t
\varphi)^2 - |{\cal D}\varphi|^2\right] + U\left[|\dot{\bf X}|^2 -
|\nabla {\bf X}|^2\right] \right\}\, .
\ee
The dimensional reduction ansatz (\ref{redan}) is now
\be
{\cal D}_5 \varphi = m\, ,\qquad \partial_5 {\bf X} ={\bf 0}\,,
\ee
and this yields the (4+1)-dimensional action
\be
S_5[\varphi,{\bf X}]= {1\over2}\int dtd^4x\, \left\{ U^{-1}\left[({\cal D}_t
\varphi)^2 - |{\cal D}\varphi|^2\right] + U\left[|\dot{\bf X}|^2 -
|\partial {\bf X}|^2\right] - m^2 U^{-1}\right\}.
\ee
Note that the potential $m^2U^{-1}$ is non-negative and has zeros at the
singularities of $U$; i.e., at the centres of the metric.

\section{Lumps and Q-lumps, Kinks and Q-kinks}
\label{sec-Klump}

As a prelude to our results on intersecting sigma-model solitons we
will briefly review some aspects of the solitons themselves. We start with
the lump and Q-lump solitons of (2+1)-dimensional HK sigma models. The
energy functional is
\be
E = {1\over2}\int d^2x \left\{ |\dot X|^2 + |\partial_1 X|^2 + |\partial_2
X|^2 + m^2 |k|^2\right\}\, .
\ee
For $m=0$ this is a massless sigma model while for $m\ne0$ it is a
massive one.  Introducing a constant unit 3-vector ${\bf n}$, we may
rewrite the energy functional as \cite{LA}
\be
E= {1\over2}\int d^2x \left\{|\dot X \mp mk|^2 + |\partial_1X - {\bf
n}\cdot {\bf I}\, \partial_2X|^2 \right\} \pm m Q + {\bf n}\cdot
{\bf L}
\ee
where $Q$ is the $U(1)$ Noether charge associated to the symmetry
generated by $k$,
\be
Q= \int d^4x\, \dot X \cdot k\, ,
\ee
and ${\bf L}$ is the topological `lump' charge
\be\label{lump}
{\bf L} = \int_{\bC} f^*(\bfO)\, ,
\ee
where $f$ is the sigma-model map from the Euclidean 2-space, viewed as
the complex plane $\bC$, and $f^*(\bfO)$ is its pullback. For fixed $Q$
and ${\bf L}$ the energy is minimized by solutions of the equations
\be
\dot X^I = \pm mk^I
\ee
and
\be\label{bpslump}
\partial_1X^I = ({\bf n}\cdot {\bf I})^I{}_J\, \partial_2 X^J\, .
\ee
The energy of solutions to these equations is given by
\be
E= |{\bf L}| + m|Q|\, .
\ee
For $m=0$ the sigma-model fields define a time-independent holomorphic map
with respect to the complex structure ${\bf n}\cdot {\bf I}$. Given a
2-cycle of the HK manifold that is holomorphic with respect to this
complex structure, there exist finite energy holomorphic maps
from $\bC$ to it. These are the sigma-model lump (or multi-lump)
solitons; it can be shown that they preserve 1/2 supersymmetry. For
$m\ne0$, the holomorphic map is time-dependent but the energy density
remains time-independent, so we now have non-static but {\it stationary}
solitons. These are the Q-lumps, which can be shown to preserve 1/4
supersymmetry.

The simplest (HK) example of both lumps and Q-lumps is
provided by a metric of the type described in the previous section with
\be\label{2cu}
U= {1\over2}\left[{1\over |{\bf X} - {\bf a}|} + {1\over |{\bf X} + {\bf
a}|}\right]
\ee
for unit vector ${\bf a}$.  Because $U$ preserves an $SO(2)$ subgroup of
the $SO(3)$ rotation group acting on the 3-vector ${\bf X}$ there is a
consistent truncation to a Kahler sigma model with $S^2$ target
space parametrized by $(\varphi, A \equiv {\bf a}\cdot {\bf X})$ with
$|A|\le1$. For this truncated model,
\be\label{simplepot}
U^{-1} = 1-A^2\, ,
\ee
and we can choose the vector potential
${\bf A}$ such that ${\cal D}\varphi= d\varphi$ \cite{GPTT}.
The two singular points at $A=\pm 1$ are the north and south
poles of the target 2-sphere, which is a holomorphic homology 2-sphere of
the original HK target space. A sigma-model lump is obtained as a solution
of the BPS equation (\ref{bpslump}) with ${\bf n}={\bf a}$. The anti-lump
is a solution of this equation for ${\bf n}=-{\bf a}$. All other
solutions, including all solutions for ${\bf n}\ne \pm {\bf a}$, are
supersymmetric `BPS flows' but not ones corresponding to {\it finite}
energy.

We now turn to the kinks and Q-kinks. These are solutions of the massive
(1+1)-dimensional sigma model, for which the energy functional is
\be
E = {1\over2}\int dx \left\{ |\dot X|^2 + |X'|^2 + m^2 |k|^2\right\}\, ,
\ee
where the prime indicates differentiation with respect to the one space
coordinate. Introducing a constant $v$ such that $v^2<1$,
we may rewrite the energy functional as \cite{AT}
\bea
E &=& {1\over2}\int dx \left\{|\dot X \mp m vk|^2 + |X' - m\sqrt{1-v^2}
({\bf n}\cdot {\bf I}) k|^2\right\} \nn
&&\ \pm\ m v Q + m\sqrt{1-v^2}({\bf n}\cdot {\bf K})
\eea
where ${\bf K}$ is the topological `kink' charge
\be
{\bf K} = \int_{\bR} f^*(i_k\bfO)\, .
\ee
For fixed $Q$ and ${\bf K}$, the energy is minimized by solutions of the
first-order equations
\be\label{kinkfl}
\dot X^I = \pm m v k^I
\ee
and
\be\label{kinkflow}
(X^I)' = m \sqrt{1-v^2} \left({\bf n}\cdot {\bf I}\right)^I{}_J\,
k^J\, .
\ee
When $v=0$ we have a static kink and when $v\ne0$ we have a stationary
Q-kink; in either case the energy is
\be
E= m\sqrt{|{\bf K}|^2 + Q^2}\, ,
\ee
and both kink and Q-kink preserve 1/2 supersymmetry.

The simplest HK examples again occur for 4-dimensional multi-centre
target spaces, for which (\ref{kinkfl})
and (\ref{kinkflow}) are equivalent to
\be\label{kinkBPS}
{\cal D}_t \varphi = \pm mv \, ,\qquad \dot {\bf X} = {\bf 0}\,
,\qquad {\bf X}' = m \sqrt{1-v^2}\, U^{-1}{\bf n}
\ee
for some unit vector ${\bf n}$. For special choices of ${\bf n}$, the
solutions of these equations will be BPS flows with finite energy;
these are the kinks and Q-kinks, which connect the vacua at the
singularities of $U$. For the 2-centre case with $U$ given by (\ref{2cu})
we can solve the equations (\ref{kinkBPS}) by again considering the
K\"ahler truncation; this shows that the finite energy BPS flows occur for
${\bf n}=\pm {\bf a}$.

In general, not every vacuum of the massive sigma model
will be connected to every other one by a kink but the vacua
must form a connected set. Note that BPS flows cannot cross for a {\it
given} ${\bf n}$,  but there is no such restriction on  BPS flows
corresponding to different choices of ${\bf n}$. For example, let ${\bf
a}$, ${\bf b}$ and ${\bf c}$ be three mutually orthogonal unit 3-vectors
and consider a 6-centre metric with
\be
U= {1\over2} \sum_\mp \left[ {1\over |{\bf X} \mp {\bf a}|} +
{1\over |{\bf X} \mp {\bf b}|} +
{1\over |{\bf X} \mp {\bf c}|} \right].
\ee
This choice preserves a discrete subgroup of $SO(3)$ that permutes the
three axes of $\bE^3$. In particular, there is a $Z_2$ subgroup that
fixes ${\bf a}\cdot {\bf X}$ but interchanges ${\bf b}\cdot {\bf X}$ and
${\bf c}\cdot {\bf X}$. This is sufficient to ensure the consistency of
the truncation to a K\"ahler sigma model with fields $(\varphi,
A \equiv{\bf a}\cdot {\bf X})$, for which
\be
U = {1\over 1-A^2} + {2\over \sqrt{1+A^2}}
\ee
Although the BPS flow interpolating between $A=-1$ and $A=1$
is now more complicated, it exists. By symmetry so
do BPS flows interpolating between ${\bf X}=-{\bf b}$ and ${\bf X}={\bf
b}$ (for ${\bf n}={\bf b}$), and between ${\bf X}=-{\bf c}$ and ${\bf
X}={\bf c}$ (for ${\bf n}={\bf c}$).

Similar considerations apply to lumps of the (2+1)-dimensional massless
sigma model; whenever there exists a massive sigma-model kink
interpolating between two centres of the $HK_4$ metric then there
exists a holomorphic map from $\bC$ to a homology
2-sphere with poles at the two centres. We will use this fact in the
following section to argue that a (4+1)-dimensional supersymmetric HK
sigma model with the 4-dimensional 6-centre target space just described
admits 1/8 supersymmetric intersections of its lump-membrane solitons.

\section{BPS equations for intersecting solitons}
\label{sec-BPSsec}

We now turn to {\it intersecting} solitons of the (4+1)-dimensional sigma
model.  The energy functional is
\be\label{energy5}
E = {1\over2}\int d^4x \left\{ |\dot X|^2 + |\bfn X|^2 +
|\partial_4 X|^2 + m^2 |k|^2\right\}
\ee
where we have set $x=({\bf x},x^4)$ and $\partial=(\bfn,\partial_4)$.
For $m\ne0$ this is a massive sigma model but we may set $m=0$ to
get the massless model. We now rewrite this energy functional as
\be
E = {1\over2}\int d^4x \left\{|\dot X \mp mk|^2 + |\partial_4 X^I - {\bf
I}^I{}_J \cdot \bfn X^J|^2\right\} \pm m Q + T
\ee
where $T$ is the surface term
\be
T= {1\over2}\int d^4x\, \bfn X^I \times \bfn X^J \cdot
\bfO_{IJ} + \int d^4x\, \partial_4 X^I \bfn X^J \cdot \bfO_{IJ}\, .
\ee
The energy is therefore minimized, for boundary conditions that fix the
value of $Q$ and $T$, by solutions of the first order `BPS' equations
\be\label{BPSt}
\dot X^I = \pm mk^I
\ee
and
\be\label{BPS}
\partial_4 X^I = {\bf I}^I{}_J \cdot \bfn X^J \, .
\ee
For sigma-model fields that are independent of $x^1$ the latter equation
reduces to
\be\label{NNSeq}
\partial_4 X^I = (I_2)^I{}_J \partial_2 X^J +
(I_3)^I{}_J\partial_3 X^J\, ,
\ee
which is equivalent to the equation considered in the context of the
(3+1)-dimensional HK sigma model in \cite{NNS}. Because the three complex
structures ${\bf I}$ obey the algebra of the quaternions, an analogous
equation holds for configurations that are independent of any one of the
four space coordinates.

When the target space is a 4-dimensional HK manifold of the type
described in the introduction, the equation ({\ref{BPSt})
becomes
\be\label{BPSt2}
\dot \varphi = \pm m\, ,
\ee
which is trivially solved, while the BPS equation ({\ref{BPS}) becomes the
set of equations
\be\label{toricBPS}
\bfn \cdot {\bf X} = U^{-1}{\cal D}_4 \varphi\, ,\qquad
\bfn\times {\bf X} = U^{-1} \bfD \varphi + \partial_4 {\bf X}\, .
\ee
For configurations that are independent of $x^4$ these equations reduce
to
\be\label{toricBPS2}
\bfn \cdot {\bf X} =0 \, ,\qquad \bfn \times {\bf X} = U^{-1} \bfD
\varphi \, ,
\ee
which describe intersecting lump-strings in the model obtained by a
trivial reduction to (3+1) spacetime dimensions. If we further specialize
to configurations that are independent of $x^3$ then we find, firstly,
that $X_1$ and $X_2$ are harmonic functions on the 12-plane; for
non-singular solutions with physical behaviour at infinity we must
therefore set $X_1=X_2=0$. We are then left with $X_3$ and $\varphi$
which are required to satisfy
\be
\partial_1 X_3 = - U^{-1}{\cal D}_2\varphi \, ,\qquad
\partial_2 X_3 =  U^{-1}{\cal D}_1\varphi
\ee
These are the equations for the sigma-model lump
in the form given in \cite{GPTT}.

We now consider the interpretation of (\ref{BPS}) or, more specifically,
(\ref{toricBPS}). Suppose that we have a solution that is asymptotically
independent of, say, $x^1$ and $x^2$ as both $x^3$ and $x^4$ become large 
(for
fixed $x^1,x^2$). Such a solution asymptotes to a lump-membrane in
the 12-plane. One can now imagine a solution of this type that is
symmetric under permutations of the four axes, in which case it will
describe the intersection of six lump-membranes, one for each of the
planes through the origin containing two of the four axes.
It is possible that 1/8 supersymmetric solutions of (\ref{toricBPS})
of this type will exist only for special models.
We will see in the following section that a simple 2-centre model allows
a special solution that can be interpreted as a non-singular intersection
of two totally orthogonal membranes. A similar type of solution must
exist for the 6-centre model described earlier with
${\bf n}={\bf a}$, for the reasons explained there, but now one can
contemplate a superposition with two similar solutions obtained from the
BPS equations for ${\bf n}={\bf b}$ and ${\bf n}={\bf c}$. By choosing
boundary conditions that preserve the permutation symmetry of ${\bf a}$,
${\bf b}$ and ${\bf c}$, and that specify the appropriate fall-off of the
fields away from any of the axes, one would expect to find a solution in
which six lump-membranes intersect in the way just described, although we
anticipate that this may be difficult to verify in practice.

\section{An explicit intersecting lump solution}
\label{sec-explicit}

We will now exhibit a particular static solution of (\ref{toricBPS}) for
the 2-centre massless HK sigma model with $U$ given
by (\ref{2cu}). We will show later that
it preserves 1/4 supersymmetry rather than 1/8 supersymmetry, so it is not
a generic solution. Nevertheless, it demonstrates that non-singular
intersections of lump-membranes can occur. We first set
\be
X_1=A\, ,\qquad X_2=X_3=0\, ,
\ee
to reduce the four equations to the two pairs of equations
\be
\label{explicit1}
\partial_1 A = U^{-1}{\cal D}_4\varphi\, ,\qquad
\partial_4 A = -U^{-1}{\cal D}_1\varphi
\ee
and
\be
\label{explicit2}
\partial_3 A = U^{-1}{\cal D}_2\varphi\, ,\qquad
\partial_2 A = -U^{-1}{\cal D}_3\varphi\, .
\ee

These equations become the kink-lump equations solved in \cite{GPTT}
on non-trivial reduction to (3+1) dimensions. As explained there,
we may now choose a gauge for which ${\cal D}\varphi =d\varphi$.
Noting that the potential $U^{-1}$ is given by (\ref{simplepot}),
we can easily solve (\ref{explicit1}) and (\ref{explicit2}). The solution is
\be
\label{lump-lump}
A= \tanh \log |Z|\, ,\qquad \varphi = {\rm arg} Z
\ee
for a function $Z(\zeta,\xi)$ that is holomorphic in the two complex
variables
\be
\zeta = x^2 + ix^3\, ,\qquad \xi = x^4 + ix^1\, .
\ee
The simplest non-trivial choice for $Z$ is
\be
\label{simple}
Z= {c\over \zeta\xi}
\ee
for some complex number $c$.  For fixed $\zeta$ we have a lump in the
$\xi$-plane and hence a membrane parallel to the $\zeta$-plane.
Similarly, for fixed $\xi$ we have a lump in the $\zeta$-plane and
hence a membrane parallel to the $\xi$-plane. Thus, the full
non-singular solution represents a pair of intersecting membranes.

Confirmation of this interpretation can be had from an investigation of
the energy. For the general solution of the form (\ref{lump-lump})
the energy density is
\be
{\cal E} = \left(1+ |Z|^2\right)^{-2} \left[ |\partial_\zeta
Z|^2 + |\partial_\xi Z|^2\right].
\ee
For the particular solution (\ref{simple}) this becomes
\be\label{ensimp}
{\cal E}(u,v) = {4|c|^2 (u + v) \over
\left(|c|^2 + uv\right)^2}
\ee
where
\be
u=|\zeta|^2\, ,\qquad v=|\xi|^2\, .
\ee
Noting that
\be
{\cal E} = -4|c|^2\left({\partial\over\partial u}
+ {\partial\over\partial v}\right)\left(|c|^2 + uv\right)^{-1}\, ,
\ee
we can easily evaluate the total energy
\be
E= 4\pi^2 \int_0^\infty du\int_0^\infty dv\, {\cal E}(u,v) \, .
\ee
Introducing a cut-off for small $u$ and $v$, we can integrate once to get
\be
E= 4\pi^2 \left[\int_\delta^\infty dv + \int_\delta^\infty du\right]
+ {\cal O}(\delta)\,.
\ee
Taking the $\delta\rightarrow 0$ limit, we recognize this as the formula
\be
E = 4\pi A_\xi + 4\pi A_\zeta
\ee
where $A_\xi$ and $A_\zeta$ are the (infinite) areas of the $\zeta$ and
$\xi$ planes, respectively. The coefficient $4\pi$ is the lump
energy, which equals the area of the unit 2-sphere for the model under
discussion; it has an obvious interpretation as the energy per unit
area (tension) of a lump-membrane.

This sigma-model solution is reminiscent of the solution of the
D=5 supermembrane equations describing a pair of intersecting
membranes (see, for example, \cite{pkt}). In fact, the D=5 supermembrane
action provides an effective description of the lump-membrane of the D=5
supersymmetric sigma model, and so we should expect to find an effective
description of the intersecting lump-membranes as a solution of it.
However, it should be remembered that the lump-membranes are membranes
with a lump core of a definite {\it non-zero} size (although this size is
arbitrary). With this in mind, it is instructive to examine the energy
density function ${\cal E}(u,v)$ as a function of $u$ for fixed $v$.
Noting that
\be
\partial_u {\cal E} = {4|c|^2 \left(|c|^2 -v^2\right)\over (|c|^2 +
uv)^3}\, ,
\ee
we see that ${\cal E}$ is a monotonically decreasing function of $u$ for
$v>|c|$ and hence in the limit of large $v$, but for $v<|c|$ the energy
density is a monotonically {\it increasing} function of $u$. We conclude
that two membranes intersect in a region of size $\sqrt{|c|}$.

Finally, we observe that there is an analogous solution of the massive
model but with $\varphi = \pm mt$. We will see later that this solution
also preserves 1/4 supersymmetry if $\varphi=mt$, but breaks all
supersymmetry if $\varphi=-mt$, despite the time-reversal invariance
of the sigma-model equations.

\section{Supersymmetry}

The condition for a supersymmetric sigma model field configuration to
preserve some fraction of the 8 supersymmetries of the sigma-model
vacuum is particularly simple for the HK manifolds described in section
\ref{sec-two}, and can be found in \cite{GGT,GPTT}. Since all
maximally-supersymmetric sigma models in spacetime dimension (2+1) and
above can be obtained by dimensional reduction (trivial or otherwise) of
the massless (5+1)-dimensional model, we need only consider that case.
The number of supersymmetries preserved by any configuration of this
theory is the number of linearly-independent {\it constant}
$SU(2)$-Majorana-Weyl spinor solutions $\lambda$ to the
equation
\be\label{susycon}
\gamma^\mu\left[\bftau \cdot \partial_\mu {\bf X} + iU^{-1}
{\cal D}_\mu\varphi\right]\lambda =0 \, ,
\ee
where $\bftau$ is the triplet of Pauli matrices. We have supressed both
the $SO(5,1)$ spinor and $SU(2)$ indices. The chirality condition on
$\lambda$ will be taken to be
\be\label{chiral}
\Gamma^{012345}\lambda= \lambda\, .
\ee
This reduces the 16 complex components of $\lambda$ to 8 complex
components but only 8 real linear combinations are linearly independent
because of the $SU(2)$-Majorana condition (which we do not give here as
we shall never need to use its explicit form). The fraction of
supersymmetry preserved by configurations of lower-dimensional sigma
models can also be determined from (\ref{susycon}) by lifting them to
(5+1) dimensions via the reduction ansatz (\ref{redan}) (with $m=0$ for
massless models).

We shall now determine the fraction of supersymmetry
preserved by a generic solution of equations (\ref{BPSt2}) and
(\ref{toricBPS}).  Using (\ref{toricBPS}) to eliminate the space
derivatives of $\varphi$ in the supersymmetry preservation condition
(\ref{susycon}), we find for $m=0$ that
\be\label{epcon1}
(\partial_j X_k)\left[ \Gamma^j \tau_k + i\Gamma^i \varepsilon_{ijk} + i
\Gamma^4\delta_{jk}\right] \lambda + (\partial_4 X_i)\left[ \Gamma^4
\tau_i -i\Gamma^i\right] \lambda =0\, ,
\ee
where $\tau_i$ $(i=1,2,3)$ are the three Pauli matrices.
For $m\ne0$ we have the additional condition
\be\label{g05}
\Gamma^{05}\lambda =\pm \lambda\, ,
\ee
where the choice of sign is inherited from (\ref{BPSt2}).

We shall begin our analysis by considering the $m=0$ case. When
$\partial_4 {\bf X}$ is non-zero we must set
\be\label{susy3}
\tau_k \lambda = i\Gamma^4 \Gamma^k \lambda \qquad (k=1,2,3).
\ee
These conditions imply that
\be\label{fourg}
\Gamma^{1234}\lambda =\lambda\, .
\ee
Using (\ref{susy3}) in (\ref{epcon1}) we find that
\be
(\partial_j X_k)\left[\Gamma^4\Gamma^{jk}
-\Gamma^i\varepsilon_{ijk}\right]\lambda=0\, ,
\ee
but this is satisfied identically as a consequence of (\ref{fourg}).
As all three conditions (\ref{susy3}) are independent we deduce that the
generic ($m=0$) solution of the BPS equations (\ref{toricBPS}) preserves
1/8 supersymmetry.

Because of the chirality condition (\ref{chiral}), the constraint
(\ref{fourg}) is equivalent to
\be\label{timecase}
\Gamma^{05}\lambda=\lambda\, .
\ee
It follows that the additional constraint (\ref{g05}) that is needed
for preservation of supersymmetry when $m\ne0$ is identically satisfied
if we choose the upper sign, but is otherwise violated.
What this means is that there can be 1/8-supersymmetric intersections of
Q-lump membranes in the massive (4+1)-dimensional theory, but only if
$\dot \varphi$ is positive. Reversing the time orientation takes the
supersymmetric solution into a non-supersymmetric one. The asymmetry
arises from the choice of chirality of the D=6 spinor $\lambda$ and
the fact that this asymmetry is maintained by a non-trivial
dimensional reduction.

For the special case in which $X_2=X_3=0$, as occurs for
the particular solution of section \ref{sec-explicit},
the supersymmetry preservation condition (\ref{epcon1}) becomes
\be
\left(\Gamma^4\partial_1 X_3 -\Gamma^1\partial_4 X_3\right)\left(1-
i\Gamma^{41}\tau_1\right)\lambda +
\left(\Gamma^2\partial_3 X_3 -\Gamma^3\partial_2 X_3\right)\left(1-
i\Gamma^{23}\tau_1\right)\lambda =0.
\ee
For generic $X_3$ this implies the conditions
\be
i\Gamma^{23}\tau_1\lambda =\lambda \, ,\qquad i\Gamma^{41}\tau_1\lambda
= \lambda\, ,
\ee
which preserve 1/4 supersymmetry. Given the chirality
condition (\ref{chiral}) these conditions imply (\ref{timecase}), so the
stationary, charged, version of the solution
of section \ref{sec-explicit} either preserves 1/4 supersymmetry
or breaks all supersymmetries, again depending on the time
orientation.

Let us now consider the simpler set of BPS equations
(\ref{toricBPS2}), appropriate to solutions that are independent of
$x^4$. In this case we need not impose the conditions (\ref{susy3}).
Instead, we may impose the conditions
\be\label{susy32}
i\Gamma_{045} \Gamma^k\tau_k \lambda = \lambda\, .
\ee
This yields
\be
(\partial_j X_k)\left[ \Gamma^{045}\Gamma^{jk} +
\Gamma^i\varepsilon_{ijk}\right]\lambda =0\, ,
\ee
but this is automatically satisfied as a consequence of the chirality
condition (\ref{chiral}).
The three new conditions (\ref{susy32}) are no longer independent
because, given the chirality condition, any two imply the third. Thus,
generic solutions of (\ref{NNSeq})
preserve 1/4 supersymmetry, in agreement with the result of \cite{NNS}.
However, (\ref{g05}) is no longer implied by these conditions so we
conclude (again in agreement with \cite{NNS}) that the time-dependent
Q-lump version of the 1/4 supersymmetric intersection of sigma-model
lump-strings preserves only 1/8 supersymmetry, but does so
irrespective of the time orientation.

\section{Calibrated M5-branes}

Consider an M5-brane in a $HK_4$ background, as described by the array
\be\label{array1}
\ba{lcccccccccc}
HK_4: & - & - & - & - & - & 6 & 7 & 8 & {\underline 9} & - \\
M5:  & 1 & 2 & 3 & 4 & 5 & - & - & - & - & - \\
\ea
\ee
The coordinate $x^9$ is identified with the angular coordinate
$\varphi$ of the $HK_4$ manifold, and the $6,7,8$ directions with the
cartesian coordinates ${\bf X}$ on $\bE^3$. If the M5-brane is fixed in
the remaining tenth direction and has vanishing worldvolume gauge
fields then its dynamics will be described at low energy by a
supersymmetric (5+1)-dimensional sigma model with $HK_4$ as its target
space.  As explained earlier, each of the singularities of the function
$U$ on $\bE^3$ that determines the $HK_4$ metric will be connected to at
least one other singularity by a holomorphic 2-sphere.  For simplicity,
suppose that there are only two singularities, separated in the
$6$-direction. Then we may wrap another M5-brane on this 2-sphere. This
second M5-brane may smoothly intersect the first one in a 3-brane,
yielding a configuration that preserves 1/8 of the 32 supersymmetries of
the M-theory vacuum. This configuration is described by the array
\be\label{array2}
\ba{lcccccccccc}
HK_4: & - & - & - & - & - & 6 & 7 & 8 & {\underline 9} & - \\
M5:  & 1 & 2 & 3 & 4 & 5 & - & - & - & - & - \\
M5:  & - & - & 3 & 4 & 5 & 6 & - & - & 9 & -
\ea
\ee
Such 3-brane intersections of M5-branes are 1/2 supersymmetric lump
solitons of the effective HK sigma model on the first
M5-brane \cite{BT,GPTT}.
An M5-brane configuration preserving only 1/16 of the supersymmetry of
the M-theory vacuum may be obtained by the addition of another
M5-brane, according to the array
\be\label{array3}
\ba{lcccccccccc}
HK_4: & - & - & - & - & - & 6 & 7 & 8 & {\underline 9} & - \\
M5:  & 1 & 2 & 3 & 4 & 5 & - & - & - & - & - \\
M5:  & 1 & - & - & 4 & 5 & 6 & - & - & 9 & - \\
M5:  & - & 2 & 3 & - & 5 & 6 & - & - & 9 & - \\
\ea
\ee
As the 5 direction is common to all M5-branes we can trivially
compactify it to arrive at an effective (4+1)-dimensional sigma model.
The array can then be interpreted as the 1/4 supersymmetric
intersection of two lump-membranes. As we saw in section
\ref{sec-explicit}, this
configuration needs only the simplest 2-centre $HK_4$ metric for its
realization.

Now consider the six-centre metric of section \ref{sec-Klump}. We argued in
section \ref{sec-BPSsec} that this would allow two
further pairs of intersecting
lump-membranes, leading to a configuration in which six lump-membranes,
in the six planes containing two axes, intersect. This possibility is
associated to the array
\be\label{array4}
\ba{lcccccccccc}
HK_4: & - & - & - & - & - & 6 & 7 & 8 & {\underline 9} & - \\
M5:  & 1 & 2 & 3 & 4 & 5 & - & - & - & - & - \\
M5:  & 1 & - & - & 4 & 5 & 6 & - & - & 9 & - \\
M5:  & - & 2 & - & 4 & 5 & - & 7 & - & 9 & - \\
M5:  & - & - & 3 & 4 & 5 & - & - & 8 & 9 & -  \\
M5:  & 1 & - & 3 & - & 5 & - & 7 & - & 9 & - \\
M5:  & - & 2 & 3 & - & 5 & 6 & - & - & 9 & - \\
M5:  & 1 & 2 & - & - & 5 & - & - & 8 & 9 & -
\ea
\ee
Note that this contains the previous array.
Also contained as a sub-array of (\ref{array4}) is
\be\label{array5}
\ba{lcccccccccc}
HK_4: & - & - & - & - & - & 6 & 7 & 8 & {\underline 9} & - \\
M5:  & 1 & 2 & 3 & 4 & 5 & - & - & - & - & - \\
M5:  & 1 & - & - & 4 & 5 & 6 & - & - & 9 & - \\
M5:  & - & 2 & - & 4 & 5 & - & 7 & - & 9 & - \\
M5:  & - & - & 3 & 4 & 5 & - & - & 8 & 9 & -
\ea
\ee
As the $4$ and $5$ directions are now common to all M5-branes we can
trivially compactify them to arrive at a configuration of three
intersecting lump-strings of a (3+1)-dimensional sigma model.

Each of the above arrays can be interpreted as describing a single
calibrated M5-brane \cite{GP,GLW}. The array (\ref{array2}) corresponds to
the K\"ahler calibration discussed in detail in \cite{GPT}. The array
(\ref{array3}) is another K\"ahler calibration, which was associated to
the kink-lump in \cite{GPTT} but has a more direct interpretation in
terms of the intersecting lump-membrane solution presented in
section \ref{sec-explicit}. The remaining arrays correspond to
exceptional calibrations. Specifically, (\ref{array4}) corresponds to a
Cayley calibrated 4-surface in $\bE^4\times HK_4$, while the sub-array
(\ref{array5}) corresponds to an associative 3-surface in
$\bE^3\times HK_4$. We need discuss only the Cayley calibration in detail
as it contains the other cases as sub-cases. We shall see below that the
low-energy limit of the equations for a Cayley-calibrated
4-surface in $\bE^4\times HK_4$ yields equations that are equivalent to
the sigma-model BPS equation (\ref{BPS}).

\section{M-theory supersymmetry}

For any of the calibrations of interest here, the equations that
govern the calibrated surface can be obtained from the requirement of
partial preservation of supersymmetry by an M5-brane.
The number of supersymmetries preserved is the dimension of the space of
solutions, for the 32-component covariantly-constant real spinor $\e$, of
the condition
\be
\label{susy1}
\G\e = \e \,,
\ee
where $\G$ is the `$\kappa$-symmetry' matrix of the M5-brane.
Let us take the $D=11$ spacetime coordinates to be
$X^M= (X^m, X^\natural)$ where $X^m=(X^\mu,X^I)$ are D=10 spacetime
coordinates for $\bE^{(1,5)}\times HK_4$. We denote the corresponding
D=11 Dirac matrices by
\be
\G_M=(\G_\mu,\G_I, \G_\star \G_{6789})\,,
\ee
where
\be
\G_\star = \G_{012345}\, .
\ee
For the chosen background we have
\be
\lbrace \G_I,\G_J \rbrace = 2 G_{IJ}
\ee
where $G_{IJ}$ is the $HK_4$ metric. The covariantly-constant spinors in
this background satisfy
\be\label{ep1}
\G_{6789} \epsilon = \epsilon
\ee
and take the form $\epsilon=f\epsilon_0$ for some non-zero
function $f$ and constant spinor $\epsilon_0$. As the function $f$
cancels we may suppose that the spinor $\epsilon$ satisfying
(\ref{susy1}) and (\ref{ep1}) is constant.

The worldvolume fields of the M5-brane are the maps $X^M(x)$ from the
worldvolume, with coordinates $(x^\mu,x^5)$ to the D=11 spacetime, and a
3-form self-dual field strength. We set the 3-form field strength to zero
and then take $X^\natural$ to be constant; this reduces the worldvolume
field content to that of a scalar multiplet of (1,0) (5+1)-dimensional
supersymmetry, and the matrix $\G$ to the $\kappa$-symmetry matrix of the
D=10 N=1 super-5-brane, for which
\be
6! \sqrt{-\det g}\, \G = \varepsilon^{\mu\nu\rho\sigma\lambda\eta}
\partial_\mu X^m\partial_\nu X^n\partial_\rho X^p \partial_\sigma X^q
\partial_\lambda X^r \partial_\eta X^s \G_{mnpqrs}\, .
\ee
In the physical gauge $X^\mu=x^\mu$, the
worldvolume metric is
\be
g_{\mu\nu} = \eta_{\mu\nu} + \partial_\mu X^m \partial_\nu X^n G_{mn}\,,
\ee
and the supersymmetry preserving condition (\ref{susy1}) becomes
\be
\label{susy2}
(\sqrt{-\det g}) \e = \biggl( 1 - \G^\mu \partial_\mu X^I \G_I -
\frac{1}{2} \G^{\mu\nu} \partial_\mu X^I \partial_\nu X^J \G_{IJ} +
\dots  \biggl) \G_\star  \e\,.
\ee
At zeroth order we have
\be
\label{zeroth}
\G_\star \e = \e \, ,
\ee
which implies that the vacuum state of the 5-brane is a $1/2$
supersymmetric M-theory configuration. Because of the condition
(\ref{ep1}), the background reduces this fraction to $1/4$; the
surviving 8 supersymmetries are those of the sigma-model vacuum.
At first order we have
\be
\label{first}
\G^\mu \partial_\mu X^I\G_I \e = 0\,.
\ee
In the special case of K\"ahler calibrations, this condition actually
implies the full M5-brane supersymmetry condition (\ref{susy2})
\cite{BT,GPTT}. Although this will not be true for calibrations in
general, we might expect (\ref{first}) to be equivalent to the
sigma model supersymmetry preservation condition (\ref{susycon}).
We shall now verify this.

Introducing a vierbein $e_I{}^A$ for the $HK_4$ metric, we may write
\be
\G_I = e_I{}^A \G_A
\ee
for flat space Dirac matrices $\G_A=(\G_i,\G_\varphi)$. For the
$HK_4$ metric (\ref{HK4}) the condition (\ref{first}) then becomes
\be
\Gamma^\mu \left[\partial_\mu X^i \G_i + U^{-1} {\cal D}_\mu \varphi
\G_\varphi\right]\epsilon =0.
\ee
This is equivalent to
\be
\Gamma^\mu \left[ \partial_\mu {\bf X} \cdot \bfsig + i U^{-1}{\cal
D}_\mu\varphi\right]\epsilon =0\, ,
\ee
where we have defined
\be
\bfsig = i\bfG \G_\varphi\, .
\ee
Note that, as a consequence of the condition (\ref{ep1}) imposed by the
HK background, these matrices obey the algebra of the quaternions, exactly
as do the Pauli matrices $\bftau$ appearing in the sigma model
preservation condition (\ref{susycon}). Since the matrices
$\Gamma^\mu$ obey the same algebra as the D=6 Dirac matrices $\gamma^\mu$
of (\ref{susycon}), the implications of (\ref{susycon}) are the same
as those of the linearized
supersymmetry preservation condition (\ref{first}), given the
additional constraints (\ref{ep1}) and (\ref{zeroth}) on the D=11
spinor $\epsilon$.

It remains for us to confirm that the BPS equation (\ref{BPS}) governing
the 1/8 supersymmetric intersections of static sigma-model solitons is
what is obtained by linearization of the M-theory supersymmetry condition
for a Cayley-calibrated M5-brane. For a Cayley calibration in $\bE^8$ we
have \cite{HL,GLW}, in our conventions
\be
(\bfn \cdot {\bf X} - \partial_4 \varphi) + {\bf i} \cdot \left( \bfn \times
{\bf X} - \bfn \varphi - \partial_4 {\bf X}\right) = (dX)^3\ terms
\ee
where ${\bf i}$ are the imaginary units of the quaternions.  A
solution of these equations can be found in \cite{R}. When the
$\bE^8$ background is replaced by $\bE^4\times HK_4$, the equation becomes
\be\label{cayley}
(\bfn \cdot {\bf X} - U^{-1}D_4 \varphi) + {\bf i} \cdot \left( \bfn
\times {\bf X} - U^{-1}\bfD \varphi - \partial_4 {\bf X}\right) =
(dX)^3\ terms
\ee
Linearizing in derivatives yields precisely the BPS equation (\ref{BPS}).

\section{Discussion}

In this paper we have found a novel BPS equation for intersecting
solitonic membranes (lumps) of (4+1)-dimensional supersymmetric HK sigma
models, for which the generic solution preserves 1/8 supersymmetry. This
equation applies to sigma models with HK target spaces of arbitrary
quaternionic dimension $n$, but the $n=1$ is special because in this case
the BPS equation can also be deduced by linearization (in derivatives) of
the equations governing Cayley-calibrated 4-surfaces in $\bE^4\times
HK_4$, and these calibrations have a physical realization in terms of an
M-theory M5-brane. The degeneration to an associative 3-surface in
$\bE^3\times HK_4$ yields the equation found in \cite{NNS} for
intersecting lump-strings of the (3+1)-dimensional HK sigma model.

We have also shown how a similar result applies to intersecting Q-lumps
of the massive (4+1)-dimensional sigma-model. A curious feature of this
case is that a supersymmetric solution is transformed into a
non-supersymmetric one by time reversal!  We should stress here that
this phenomenon is not just a feature of generic solutions that preserve
1/8 supersymmetry (which remain hypothetical) but is also a feature of an
explicit special solution that preserves 1/4 supersymmetry.
Although this may seem a strange phenomenon, we
remind the reader that a similar phenomenon has long been known for space
reversal \cite{KK}.

We have certainly not exhausted the possible sigma-model soliton
intersections in this paper. New possibilities, preserving only 1/8
supersymmetry, are likely to occur in models for which the HK target 
space has a quaternionic dimension $n\ge1$, as shown by the 
multi-domain walls of \cite{GTT2} and the intersecting domain 
walls of \cite{GTT}, although any M-theory interpretation of these new
possibilities will have to be rather different from the $n=1$ cases
discussed here. However, even for $n=1$ there are likely to be further
possibilities; we will conclude with a brief mention of one
such case. Let us write the energy functional (\ref{energy5}) as
\be
E = {1\over 2}\int d^4x\, \left\{ |\dot X -\sigma mvk|^2 + |{\bf
I}\cdot\bfn X -m\eta \sqrt{1-v^2}\, k|^2\right\} + \sigma mvQ + \eta m K
\ee
where $\sigma$ and $\eta$ are two signs, and $K$ is the topological charge
\be
K = \int d^4x\, (i_k \bfO)_I\cdot \bfn X^I\, .
\ee
The energy is minimized, for fixed $Q$ and $K$ by solutions of the
first-order equations
\be
\dot X= \sigma mv k\, ,\qquad
{\bf I}\cdot \bfn X = \eta m\sqrt{1-v^2}\, k\, .
\ee
This equation is solved by the 1/4-supersymmetric Q-kink-lump of
\cite{GPTT} but generic solutions preserve 1/8 supersymmetry.

\acknowledgments

It is a pleasure to thank Jerome Gauntlett for helpful discussions. R.P.
thanks Trinity College for financial support.

%%%%%%%%%%%%%%%%%%%%%%%%%%%%%%%%%%%%%%%%%%%%%%%%%%%%%%%%%%%%%%%%%%%%%%%%%%%%
%%
%%%%%%%%%%%%%%%%%%%%%%%%%%%%%%%%%%%%%%%%%%%%%%%%%%%%%%%%%%%%%%%%%%%%%%%%%%%%
%%
%%%%%%%%%%%%%%%%%%%%%% BIBLIOGRAPHY
%%%%%%%%%%%%%%%%%%%%%%%%%%%%%%%%%%%%%%%%%%
%%%%%%%%%%%%%%%%%%%%%%%%%%%%%%%%%%%%%%%%%%%%%%%%%%%%%%%%%%%%%%%%%%%%%%%%%%%%
%%
%%%%%%%%%%%%%%%%%%%%%%%%%%%%%%%%%%%%%%%%%%%%%%%%%%%%%%%%%%%%%%%%%%%%%%%%%%%%
%%
%\newpage

\end{document}